\documentclass[letter]{aa} 
\usepackage{graphicx}
\usepackage{txfonts}
\usepackage{natbib}
\bibpunct{(}{)}{;}{a}{}{,} 
\usepackage{amssymb} 
\begin{document}
   \title{Selective Principal Component Extraction and Reconstruction: A Novel Method for Ground Based Exoplanet Spectroscopy}


   \author{A. Thatte
         \inst{1}
          \and
          P. Deroo \inst{2} \and M. R. Swain \inst{2}
          }

   \institute{Woodruff School of Mechanical Engineering, MRDC Building, Room 4111, 
   Georgia Institute of Technology, Atlanta, Georgia 30332-0405, USA \\
  	  \email{azamthatte@gatech.edu}
 	   \and
             Jet Propulsion Laboratory, California Institute of Technology, 
             4800 Oak Grove Drive, Pasadena, California 91109-8099, USA\\
             }

   \date{}

 
  \abstract
{Infrared spectroscopy of primary and secondary eclipse events probes the composition of exoplanet
    atmospheres and, using space telescopes, has detected H$_2$O,
    CH$_4$ and CO$_2$ in three hot Jupiters. However, the available
    data from space telescopes has limited spectral resolution and
    does not cover the $2.4\, -\, 5.2\,\mu$m spectral region. While
    large ground based telescopes have the potential to obtain
    molecular-abundance-grade spectra for many exoplanets, realizing
    this potential requires retrieving the astrophysical signal in the presence of large Earth-atmospheric
    and instrument systematic errors.}  
{Here we report a wavelet-assisted, selective principal component extraction method for
    ground based retrieval of the dayside spectrum of HD\,189733b from
    data containing systematic errors.} 
{The method uses singular
    value decomposition and extracts those critical points of the
    Rayleigh quotient which correspond to the planet induced
    signal. The method does not require prior knowledge of the planet
    spectrum or the physical mechanisms
    causing systematic errors. } 
{The spectrum obtained with our
    method is in excellent agreement with space based measurements made
    with HST and Spitzer (Swain et al. 2009b; Charbonneau et al. 2008)
    and confirms the recent ground based measurements (Swain et
    al. 2010) including the strong $\sim 3.3\, \mu$m emission. }
   {}


\titlerunning{A Novel Method for Ground-Based Exoplanet Spectroscopy} 
\maketitle

%

\section{Introduction}

Detection of molecules in exoplanet atmospheres via infrared
spectroscopy from space-based telescopes is now routine
\citep{Swain_2008, Grillmair_2008, Swain_2009a, Swain_2009b,
  Tinetti_2010}.  Recently, \cite{Swain_2010} demonstrated molecular
spectroscopy of an exoplanet atmosphere with ground-based
measurements.  With the availability of numerous large ground based
telescopes equipped with infrared spectrometers, there is a great
potential to obtain a large quantity of ``molecular-abundance-grade''
spectra; realizing this potential requires developing optimal signal
extraction algorithms to retrieve the spectral signature of an
exoplanet atmosphere in the presence of large Earth-atmospheric and
instrument systematic errors.  Here we present a method based on
Principal Component Analysis (PCA) capable of detecting the exoplanet
emission spectrum from the ground. PCA is a well established method
with numerous astronomomical applications; it has been used to search
for an exoplanet signal in ground-based spectroscopic observations
\citep{Brown_2002} and a related method is used in SysRem for
exoplanet eclipse detections \citep{Tamuz_2005}. We apply our PCA
based method to extract the dayside emission spectrum of HD189733b in
the K and L bands and we compare the resulting spectrum with
previously reported results \citep{Swain_2010}, which were obtained
using a different method.

\section{Observations and Initial Calibration}

The spectrum presented here is based on the same observations used in
\cite{Swain_2010} but analyzed using the Selective Principal Component
Extraction and Reconstruction (SPCER) method described in Sect.~3. A
secondary eclipse of HD~189733b was observed on August
11$^{\mathrm{th}}$, 2007 using the SpeX instrument mounted on the NASA
Infrared Telescope Facility (IRTF). The spectroscopic time series
begins approximately one hour before the onset of ingress and ends
approximately one hour after the termination of egress. The details of
the observations and the reduction with SpexTool are presented in
\cite{Swain_2010}. The result of the standard SpexTool calibration is
a flux density time series with $\sim\,4\,\%$ variations. We then
employ the initial calibration step (normalization \& airmass
correction) outlined in \cite{Swain_2010}. Finally, we separate the
exoplanet eclipse astrophysical signal from the residual systematic
errors using the SPCER method and thus obtain the exoplanet emission
spectrum.


\section{Method}

The astrophysical signal from the exoplanet is present in all spectral
channels and we have developed the SPCER method to identify this
signal (i.e.~the exoplanet eclipse) and separate it from other signal
components which are not of astrophysical origin and present in all
spectral channels (i.e.~systematic errors). The method is based on
Principal Component Analysis (PCA), which performs an orthogonal
transformation that aligns the transformed axes in the directions of
maximum variance. From statistical viewpoint, the eigenvectors of the
covariance matrix of a dataset are the critical points of the Rayleigh
quotient and when ranked according to the magnitude of the
eigenvalues, they represent the axes of maximum variance of that
data. To successfully separate non-random signals (systematic errors;
the astrophysical signal) from random noise, it is desirable that the
time series data has a high signal to noise (SNR) ratio. Therefore, we
prefilter the time series prior to applying the SPCER method. After
prefiltering, we perform singular value decomposition of a set of time
series at different wavelengths. This conventiently separates the
astrophysical signal from some of the larger systematic errors present
in the data. We then reconstruct the astrophysical signal dominated
time series and measure the exoplanet emission spectrum. In this
section, we outline the methodology. In Sect.~4, we discuss the
results of applying this method to the HD~189733b dataset.

\subsection{prefiltering}
To clean the time series and at the same time retain the dynamic
features of the data, we have implemented wavelet transform based
adaptive signal extraction. 
Wavelets are time-localized because the supports of wavelet functions
are finite. This makes wavelets excellent for representing discrete
events (e.g.~abrupt changes linked to outlier points). The wavelet
basis functions are constructed from a single function, termed the
``Mother Wavelet'', $\psi_{0,0}$. The time series is represented as a
set of wavelet functions, $\psi_{j,k}$, constructed by a combination
of dilation and translation of the mother function: $\psi_{j,k} =
2^{j/2}\psi_{0,0}(2^jt-k)$.
Here, we implemented the standard wavelet
decomposition-thresholding-reconstruction procedure which is widely
used in signal and image processing fields \citep{Sidney_1998,
  Mallat_1989, Mallat_1999, Donoho_1995, Donoho_1995b}. As mother
wavelet, we use the standard Daubechies 3 wavelet
\citep{Mallat_1999}.
We obtain the wavelet decomposition by a multi-scale representation
and using the transform coefficients $\alpha_k^j$ ($k^{th}$ scaling
function at $j^{th}$ scale) and $\beta^j_k$ ($k^{th}$ wavelet function
at $j^{th}$ scale) defined as,
\begin{eqnarray}
  \alpha_k^j & = & \int f(t)\phi(2^jt-k) dt = \sum_n
  h(n)\alpha^{j+1}_{n+2k} \\
 \beta^j_k & = & \int f(t) \psi_{0,0}(2^jt-k)dt = \sum_ng(n)\alpha^{j+1}_{n+2k}
\end{eqnarray}
where the scaling function $\phi$ can be respresented as a
  signal with a low-pass spectrum and expressed in terms of wavelets
  \citep[see][]{Mallat_1989}; $f(t)$ is the time series for a
  specific wavelength. 
$h(n)$ and $g(n)$ represent low pass and high
  pass filter coefficients respectively and are obtained using
  Daubechies 3 wavelets in MATLAB. For a detailed description on the
  theory of wavelet decomposition, the reader is referred to
  \cite{Sidney_1998, Mallat_1989}. To produce a smoother estimate and
  continuous mapping during wavelet shrinkage, we implement a soft
  thresholding scheme, $\eta_T(w_j)=\textrm{sign}(w_j)(|w_j|-T)_+$,
  where $w_j$ is substituted by the different $\alpha_k^j$ and
  $\beta^j_k$, 
  along with the universal threshold $T=\sigma_N\sqrt{2\ln(N)}$,
  where N is the current level of wavelet decomposition and
    $\sigma_{{N}}$ is the approximation of noise at this level
    obtained using the robust median estimator. After
  thresholding, we reconstruct the time series using reconstruction
  filters and the iterative reconstruction method. Our wavelet based
  signal extraction algorithm accepts signal $F_{\lambda}(t)$ as
  input and returns the de-noised signal $F_{\lambda}^w(t)$. As
  a result of this procedure, we get on average a 56\,\% improvement
  in the $\frac{\textrm{standard\,deviation}}{\textrm{mean}}$ of the individual time series
  measurements.

   \begin{figure}
   \centering
   \includegraphics[width=\hsize]{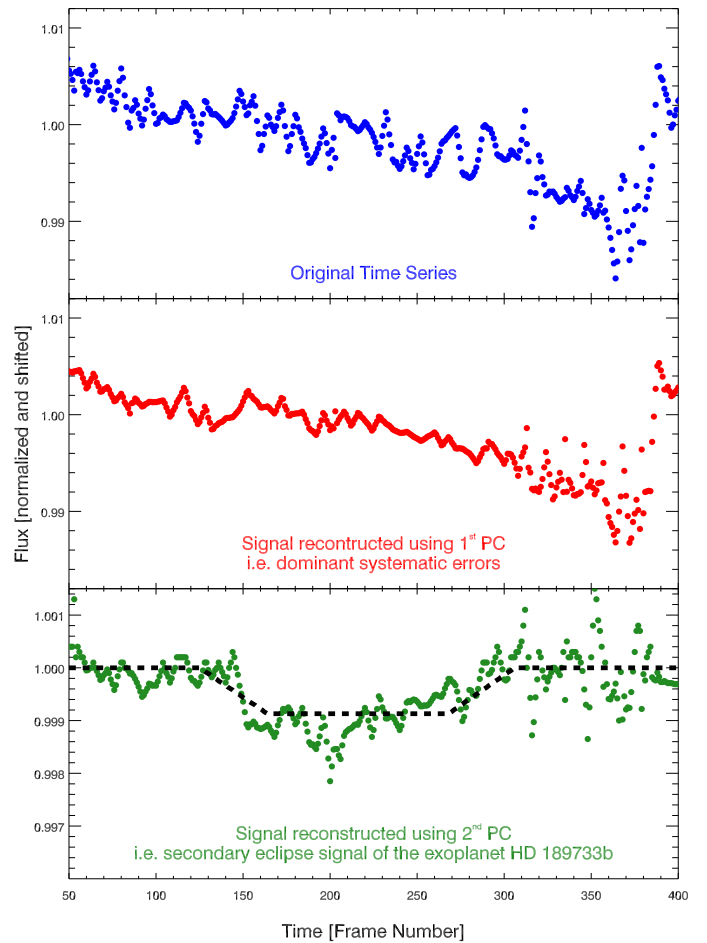}
   \caption{Separating the astrophysical signal from systematic
     errors.  The different panels depict: \textit{top:} an original,
     normalized and airmass corrected time series, \textit{middle:}
     signal reconstructed using the first PC, \textit{bottom:} signal
     reconstructed using the second PC. In the bottom panel, we show
     using a black dashed line the secondary eclipse lightcurve
     overplotted on the reconstructed data. The singular value
     decomposition of a set of time series at different wavelengths
     allows us to extract the exoplanet signature, largely free of
     systematic errors. We show the data at the original, unbinned
     time sampling. }
         \label{Fig1}
   \end{figure}

\subsection{separating multiple patterns in the data}
We start by averaging the time series $F_{\lambda}^w(t)$
  in the spectral domain (mean of 5 adjacent spectral channels) to
  reduce the number of PCs and get the time series
  $\hat{F}_{\lambda}^w(t)$. Given $X$, a subset of
  $\hat{F}_{\lambda}^w(t)$ composed of P spectral
channels, we center $X$ by subtracting $\bar{X}$, the mean of each
column of $X$. Through singular value decomposition we find the
eigenvalues ($\bar{\lambda}=\lbrace\lambda_1,\lambda_2, \dots
,\lambda_P\rbrace$) and the eigenvectors (principal components $
\bar{\bar{C}}=\lbrace C_1,C_2\dots C_P\rbrace$) of the covariance
matrix of $X$. We then project the centered data onto the principal
component axes to get $\bar{\bar{\mathfrak{R}_{pc}}}$, the
representation of X in the principal component space using
  $\bar{\bar{\mathfrak{R}_{pc}}} = X \otimes \bar{\bar{C}}$ where
  $\otimes$ denotes matrix multiplication. 


For our purposes, we want to decompose the signal present in the
ensemble of time series into various components, e.g. flux variation
due to secondary eclipse and Earth-atmospheric effects etc. For this, we
transform each individual principal component $C_r\,(C_r=\bar{\bar{C}}(r),\,r=1,2,\dots,N) $ into the time domain as, 
\begin{equation}
\bar{\bar{S}}_{r^{th}pc}=\bar{X}+\bar{\bar{\mathfrak{R}}}_{pc}(r)\otimes (\bar{\bar{C}}(r))^T
\qquad \mathrm{for }\, r=1,2,\dots,N.
\end{equation}

At this point, what we have done is transform our set of time series
so that it is expressed in terms of patterns  (principal components)
that optimally represent the covariance (commonality) of $X$. 
Each $\bar{\bar{S}}_{r^{th}pc}$ is such a reconstructed pattern and it
shows how the signal would have looked like if it was due to this
single PC. We can now determine which PCs represent the
  eclipse and which represent confusing systematic errors.

\subsection{extracting the exoplanet spectrum}

To select the principal components corresponding to the eclipse, we
calculate the linear correlation coefficient ($CC$) between each of
the individually reconstructed signals $\bar{\bar{S}}_{r^{th}pc}$ and
the expected light curve shape. No prior knowledge on the depth of the
eclipse is required for this; only the exoplanet
ephemerides\citep[][for HD\,189733]{Winn_2007} is needed. We select a
set of principal components capturing the eclipse event using
selection criteria described below and reconstruct the data using
these components ($\bar{\bar{S}}_{Rpc}$). \emph{As a result of this
  reconstruction, we get an exoplanet eclipse light curve in
    which the effect of confusing systematics is greatly reduced.}
For each wavelength channel in $X$, we determine the eclipse
  depth and the error bar based on the standard deviation of the time
  series in-eclipse and out-of-eclipse.  The eclipse depth for $X$ is
  the average eclipse depth and the error bar is composed of the error
  on the eclipse depth for each wavelength channel and the standard
  deviation of the depths for each wavelength channel in
  $X$: 

\begin{equation}
\sigma_{\textrm{eclipse\,depth}} = \sqrt{
(\frac{\sqrt{\sum_{i=1}^K(\frac{\sigma_{in}^2}{N_{in}} +
    \frac{\sigma_{out}^2}{N_{out}}})}{\sqrt{K}})^2
+ 
\sigma_{depths}^2
}
\end{equation}
with $\sigma_{in}$ and $\sigma_{out}$ the standard deviation
  of the flux in-eclipse and out-eclipse regions for each wavelength
  channel in $X$; $N_{in}$ and $N_{out}$ the number of observations in
  and out of eclipse; $K$ the number of wavelength channels in $X$;
  $\sigma_{depths}$ the standard deviation of the eclipse depth for
  the different wavelength channels in $X$. 


The selection criteria to decide whether a principal component
captures the exoplanet eclipse event are:
\begin{enumerate}
\item $|{CC}|_{PC,i}>|{CC}|_{\mathrm{threshold}}, $ with
  $|{CC}|_{\mathrm{threshold}} = $ $
  \mathrm{mean}(|CC|)+\mathrm{std\,dev}(|CC|)$; the correlation
  coefficient of the PC must be large enough, such that it describes
  well the eclipse event.

\item $\, R_{PC,i}<R_{\mathrm{threshold}}$ with $R_{PC,i}$ the
  eigenvalue rank of the i$^{\mathrm{th}}$ selected PC and 
    $R_{\mathrm{threshold}}$ is the cutoff value for rank. We have
    chosen $R_{\mathrm{threshold}} = 9$ (see later).
\item $|R_{PC,i} - R_{PC,i+1}|<\Delta_{\mathrm{threshold}}$; 
    $\Delta_{\mathrm{threshold}}$ is the cutoff value for the
    difference in the ranks of the selected PCs. We have chosen
    $\Delta_{\mathrm{threshold}} = 6$ (see later).
\end{enumerate}
If no PC is found under these selection criteria, we assign an upper
limit to the eclipse depth, matching the lowest eclipse depth found
in the dataset and conclude that we could not find the eclipse for those
wavelength channels.

  \begin{figure}
   \centering
   \includegraphics[width=\hsize]{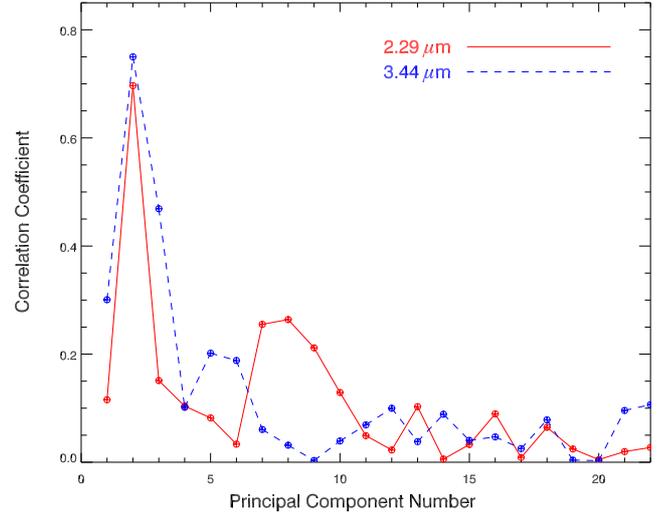}
   \caption{Selecting the principal components which capture the
     exoplanet eclipse. The linear correlation coefficients between
     the PCs and the model eclipse light curve are shown for
     representative wavelengths for the K and L-band. The eclipse is
     captured by PCs with a high correlation coefficient. The
     selection of PCs is rather straightforward with very large
     correlation coefficients being found at low PC numbers.  }
         \label{Fig2}
   \end{figure}
\begin{figure*}
\centering
\includegraphics[width=17.5cm]{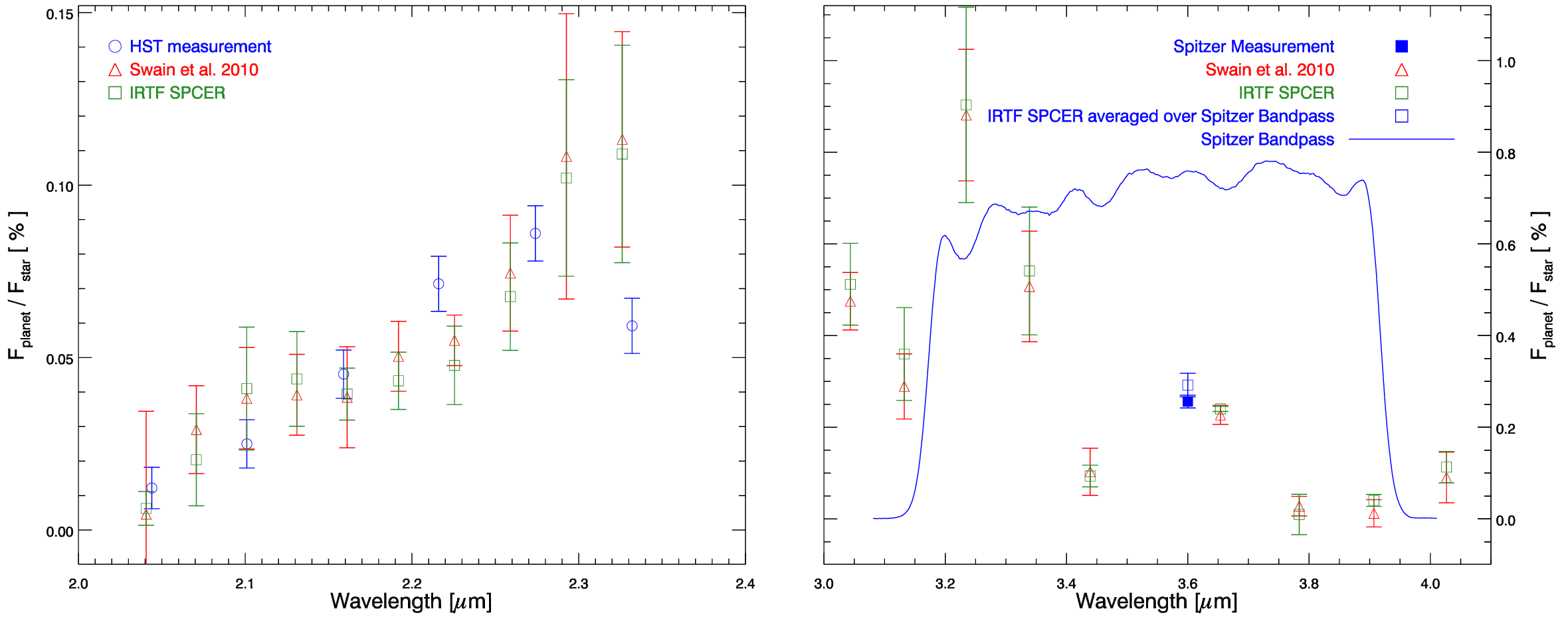}
\caption{Dayside emission spectrum. The K-band (\textit{left}) and
  L-band (\textit{right}) HD~189733b emission spectrum obtained using
  SPCER is shown using green squares and is compared to the
  space-based HST spectrum \citep{Swain_2009b} and Spitzer broad-band
  photometry \citep{Charbonneau_2008} and the ground-based spectrum
  reported in \cite{Swain_2010}. We also show the Spitzer passband
  (blue line) for the 3.6$\,\mu$m photometric point and have averaged
  the SPCER result over this passband to make comparison easy (blue
  open symbol). There is excellent agreement between the space and
  ground-based datasets. }
         \label{Fig3}
   \end{figure*}
\section{Results and Discussion}

We have applied the method outlined above to the spectroscopic time
series on the secondary eclipse of HD\,189733b observed with the
IRTF. We chose to use the same spectral binning as in
\cite{Swain_2010}, such that we can easily compare the results. In the
K-band and L-band, we used 100 and 150 spectral channels respectively
to construct $X$. We then processed the data through the different
steps outlined above and a sample result of the principal component
extraction is shown in Fig.~\ref{Fig1}. Blue symbols show the
normalized original light curve which is dominated by residual
systematic errors.  Red and green curves show the signal reconstructed
using the first and the second PC respectively. The model eclipse
curve is shown in black. The first PC can be seen to represent much of
the systematic errors present in the original signal, while the second PC
is clearly the exoplanet eclipse. This demonstrates the potential of
SPCER in separating systematic effects for ground based observations
and in retrieving the exoplanet signal. It does so without requiring
knowledge of the physical mechanism causing the corruption nor
requiring prior knowledge on the exoplanet.

To determine the exoplanet spectrum, we have selected
$\bar{\bar{S}}_{Rpc}$ using the criterium outlined above with
$R_{\mathrm{threshold}} =9 $ and $\Delta_{\mathrm{threshold}} = 6$. In
Fig.~\ref{Fig2}, we illustrate the calculated correlation coefficients
with the light curve for K and L-band wavelengths. It is clearly seen
that a PC either matches really well ($CC >0.65$) or not very well,
making the selection rather straightforward and changes in the
selection criterium have little influence. The result of the
procedure, the exoplanet spectrum of HD\,189733b, is shown in
Fig.~\ref{Fig3}. The SPCER method reproduces the measurements by HST
\citep{Swain_2009b}, the Spitzer photometry \citep{Charbonneau_2008}
and those reported in \citep{Swain_2010} extremely well, within the
error bars.

A few considerations we would like to mention are:
\begin{itemize}
\item \textit{importance of pre-cleaning the data; removing the
    biggest errors first}. The data processing outlined in
  \cite{Swain_2010} includes removal of systematic errors correlated
  in wavelength and airmass. If this step is not taken, it is still
  feasible to find the eclipse signal, but at lower eigenvalue
  rank. For instance, not performing the airmass correction outlined
  in \cite{Swain_2010}, shifts the eclipse signal from a typical
  eigenvalue rank 2 to eigenvalue rank 4.
\item \textit{importance of selecting (sometimes) multiple PCs}. There
  is a lack of control in the choice of orthogonal basis functions in
  the singular value decomposition. This means that the eclipse is not
  always captured in a single PC, and several PCs (typically two) may
  be needed to represent the eclipse. When the reconstruction is done
  based solely on one PC, the spectral shape is preserved, but the
  average eclipse depth can change by up to $\sim 20\,\%$.
\item \textit{possible improvement by post-processing the PCs}. The
  SPCER method does not necessarily decompose the data entirely into
  an eclipse and non-eclipse component, because no priors are included
  (which is also the strength of the method). As such, some residual
  systematic error can be convolved with the eclipse signal. Simple
  post-processing, like fitting a polynomial to the out-of-eclipse
  part of the light curve, can potentially enhance the SNR of the
  exoplanet spectrum. For this particular dataset, this was not
  necessary, but might prove advantageous for other datasets.
\item \textit{further improvements are likely possible by
      incorporating the known eclipse shape}.  In the method presented
    here, no prior knowledge of the eclipse shape is used during the
    disentanglement of systematic effects and eclipse signal. Methods
    that incorporate priors, such as an iterative matched filter, have
    the potential to improve the method presented here. 
\item \textit{equal signal to noise ratio for the different
      wavelength channels is assumed when using PCA}. For the dataset
    analyzed here, the different wavelength channels in each spectral
    bin $X$ differ only marginally in count rate (the standard
    deviation of the normalized SNR is on average 4\,\% and is always
    less than 13\,\% for the different wavelength bins $X$) and regular PCA is therefore
    appropriate. If this is not the case, a SysRem type of
    down-weighting of wavelength channels with low count rates
    \citep{Tamuz_2005}, is needed. 
\end{itemize}

   In summary, the coupled wavelet denoising-SPCER method presented
   here is a new and powerful method for ground based exoplanet
   calibration. A useful aspect of the method lies in its ability to
   reject systematic errors without the need for knowledge of the
   underlying physical mechanism. This gives us the ability to
   separate the original planetary signal from the relatively large
   systematic effects.  The results of the SPCER-based calibration for
   the emission spectrum of HD\,189733b are in excellent agreement
   with those obtained with HST \citep{Swain_2009a} and Spitzer
   \citep{Charbonneau_2008} and confirm the recent ground based
   measurements \citep{Swain_2010}. The strength of SPCER in
   retrieving the exoplanet spectrum from ground based observations is
   a proof of concept for its more general application in fields where
   signal deeply buried under noise must be extracted; the
   method has applications in a variety of fields including earth and
   atmospheric sciences, telecommunication systems, measurement
   instruments, biomedical engineering, optics, image processing and
   controls, where problems of not having good signal to noise ratio
   before signal amplification are prevalent or where pattern
   recognition is critical.

\bibliographystyle{aa}

\end{document}